\documentclass[sigconf]{acmart}

\usepackage[T1]{fontenc}
\usepackage{xcolor}
\usepackage{xcolor-material} %
\usepackage{makecell} %
\usepackage{rotating} %
\usepackage{multirow} %
\usepackage{siunitx}
\usepackage{transparent}
\usepackage{url}
\usepackage[hyphenbreaks]{breakurl}
\usepackage{hyperref}

\usepackage{tabularx}
\usepackage{graphicx}
\usepackage{acro}
\usepackage{booktabs}
\usepackage{flushend}
\usepackage{enumitem}
\usepackage{tcolorbox}

\newtcolorbox{takeaway}{
  left=0.5mm,
  right=0.5mm,
  bottom=0.5mm,
  top=0.5mm,
  before upper={\textit{Take Away: }}
}
\newcommand{\name}{SWICS} %
\newcommand{\goodChannel}{5G-GC} 
\newcommand{\badChannel}{5G-DC} 

\newcommand{\importantnote}[3]{\textbf{\textcolor{#2}{#1: #3}}}

\newcommand{\stl}[1]{\importantnote{SL}{cyan!75}{#1}}
\newcommand{\mh}[1]{\importantnote{MH}{brown!75}{#1}}
\newcommand{\sm}[1]{\importantnote{SM}{blue!75}{#1}}

\newcommand{\added}[1]{{\color{blue}{#1}}}

\definecolor{myYellow}{HTML}{FFF59D}  %
\definecolor{myOrange}{HTML}{FFCC80}  %
\definecolor{myGreen}{HTML}{82C90A}
\definecolor{forestgreen}{RGB}{150,210,100}
\definecolor{transparentRed}{HTML}{ff7f7f}
\definecolor{transparentBlue}{HTML}{7f7fff}
\definecolor{transparentGreen}{HTML}{bfe384}
\definecolor{alertBlue}{HTML}{510ac9}
\definecolor{alertRed}{HTML}{510ac9}
\definecolor{softred}{rgb}{1.8,0.3,0.2}

\renewcommand{\mh}[1]{}
\renewcommand{\stl}[1]{}
\renewcommand{\sm}[1]{}
\renewcommand{\added}[1]{}

\DeclareAcronym{TCP}{short=TCP, long=Transmission Control Protocol}
\DeclareAcronym{IP}{short=IP, long = Internet Protocol}

\DeclareAcronym{ICS}{short=ICS, long = Industrial Control System}
\DeclareAcronym{PLC}{short=PLC, long = Programmable Logic Controller}
\DeclareAcronym{SCADA}{short=SCADA, long=Supervisory Control and Data Acquisition}
\DeclareAcronym{IT}{short=IT, long=Information Technology}
\DeclareAcronym{OT}{short=OT, long=Operational Techology
}
\DeclareAcronym{HMI}{short=HMI, long=Human-Machine Interface}
\DeclareAcronym{DMZ}{short=DMZ, long= demiliterized zone}
\DeclareAcronym{IIoT}{short=IIoT, long = Industrial Internet of Things}

\DeclareAcronym{DES}{short=DES, long=Discrete Event Simulation}

\DeclareAcronym{IIDS}{short=IIDS, long= Industrial Intrusion Detection System}
\DeclareAcronym{IDS}{short=IDS, long=Intrusion Detection System}
\DeclareAcronym{IAT}{short=IAT, long=inter-arrival time}

\DeclareAcronym{URLLC}{short=URLLC, long=ultra-low latency communication}
\DeclareAcronym{EMBB}{short=eMBB, long=Enhanced Mobile Broadband}
\DeclareAcronym{MMTC}{short=MMTC, long=Massive Machine-Type Communications}
\DeclareAcronym{UE}{short=UE, long= user equipment, plural-form=user equipment}
\DeclareAcronym{gNB}{short=gNB, long=  gNodeB}
\DeclareAcronym{IOT}{short=IoT, long=Internet-of-Things}

\DeclareAcronym{MITM}{short = MitM, long = Man-in-the-Middle}
\DeclareAcronym{DOS}{short = DoS, long = Denial of Service}

\DeclareAcronym{RSSI}{short=RSSI, long=Received Signal Strength Indicator}

\begin{document}

\title{
    Security Implications of 5G Communication in Industrial Systems
}

\author{Stefan Lenz}
\orcid{1234-4564-1234-4565}
\email{lenz@spice.rwth-aachen.de}
\authornotemark[1]
\affiliation{
    \institution{RWTH Aachen University}
    \city{Aachen}
    \country{Germany}
}

\author{Sotiris Michaelides}
\email{michaelides@spice.rwth-aachen.de}
\authornote{Both authors contributed equally to this research.}
\orcid{1234-5678-9012}
\affiliation{%
  \institution{RWTH Aachen University}
  \city{Aachen}
  \country{Germany}
}

\author{Moritz Rickert}
\email{moritz.rickert@rwth-aachen.de}
\affiliation{
    \institution{RWTH Aachen University}
    \city{Aachen}
    \country{Germany}
}

\author{Jonas Holtwick}
\email{jonas.holtwick@rwth-aachen.de}
\affiliation{
    \institution{RWTH Aachen University}
    \city{Aachen}
    \country{Germany}
}

\author{Martin Henze}
\email{henze@spice.rwth-aachen.de}
\orcid{1234-4564-1234-4565}
\affiliation{
        \institution{RWTH Aachen University} 
        \city{Aachen}
        \country{Germany}
}
\additionalaffiliation{
    \institution{Fraunhofer FKIE}
    \city{Wachtberg}
    \country{Germany}
}

\copyrightyear{2026}
\acmYear{2026}
\setcopyright{cc}
\setcctype{by}
\acmConference[CPSS '26]{The 11th ACM Cyber-Physical System Security Workshop}{June 01--05, 2026}{Bangalore, India}
\acmBooktitle{The 11th ACM Cyber-Physical System Security Workshop (CPSS '26), June 01--05, 2026, Bangalore, India}
\acmDOI{10.1145/3775042.3807886}
\acmISBN{979-8-4007-2313-1/2026/06}
\begin{abstract}
Traditionally, industrial control systems (ICS) were designed without security in mind, prioritizing availability and real-time communication.
As these systems increasingly become targets of powerful adversaries, security can no longer be neglected.
Driven by flexibility and automation needs, ICS are transitioning from wired to 5G communication, introducing new attack surfaces and a less reliable communication medium, thereby exacerbating existing security challenges.
Given their critical role in society, a comprehensive evaluation of their security is imperative.
To this end, we introduce \name{}, a fully virtual testbed simulating an ICS in a realistic 5G environment, and study how this transition affects security under varying channel conditions.
Our results show three key findings: under optimal channel conditions, industrial 5G networks can achieve resilience comparable to wired systems, while degraded channel conditions can amplify traditional attacks, threaten system stability, and undermine detection mechanisms based on predictable traffic patterns.
We further demonstrate the inherent limits of securing 5G channels for ICS through eavesdropping and jamming on the open-air interface.
Our work highlights the interplay between security and 5G channel conditions, showing that traditional security controls may no longer be sufficient and motivating further research.

\end{abstract}



\maketitle%

\section{Introduction}

Industry~4.0 and the advent of the Industrial Internet of Things have revolutionized \acp{ICS}, which manage physical processes in manufacturing facilities and critical infrastructure~\cite{henze2017network}.
Traditionally isolated, these control systems have evolved into highly interconnected networks~\cite{knapp2024_CH3}.
In addition, security has historically been deprioritized in favor of high availability and reliable real-time communication~\cite{knapp2024_CH3}.
However, recent cyberattacks against such systems\textemdash{}e.g., Stuxnet and the attacks on the Ukrainian power grid~\cite{attacks_report}\textemdash{}have shown that security can no longer be neglected to prevent harm of humans and infrastructure~\cite{dehlaghi2023icssim, conti2021survey, wolsing2022ipal}.

At the same time, \acp{ICS} are undergoing a second major transformation, as rising demands for flexibility and automation drive a shift from wired to wireless communication~\cite{aijaz2020private5G}.
While industrial companies have long explored wireless approaches, earlier technologies could not meet stringent operational requirements~\cite{michaelides2025industry5G}.
Recent advances\textemdash{}most notably 5G with sub-1~ms latencies—make wireless solutions viable for even the most demanding \acp{ICS}~\cite{5gacia_use_cases,5gaciaEdgeindustrial}. 
As a result, this transition is increasingly materializing in practice~\cite{aijaz2020private5G, michaelides2025industry5G}.
However, despite these advantages, 5G networks complicate \ac{ICS} security. 
Communication is now subject to physical-layer effects such as interference and propagation loss, which can impact process stability, weaken existing security controls including isolation strategies and introduce new attack vectors. 
Consequently, securing such systems has become more challenging than ever, especially as they are increasingly targeted by sophisticated adversaries~\cite{securityprivate,michaelides2025industry5G,attacks_report}.

To stay ahead of these threats, security must be considered from the outset when designing the next generation of industrial 5G-enabled \acp{ICS}, requiring a holistic understanding of how 5G communication affects security, including the effectiveness of existing security controls and the emergence of new attack vectors.
To lay the foundation for these efforts, in this paper we provide the first comprehensive study of the security implications of transitioning to 5G in \acp{ICS} by (i) evaluating the process stability under attacks, (ii) assessing the effectiveness of existing security measures, and (iii) investigating additional attack vectors on the physical process.
This work aims to raise awareness of the security implications of 5G-enabled ICS and stimulate further research toward proactive security design. More specifically, our contributions are:
\begin{enumerate}[leftmargin=*,topsep=0em]
    \item We develop and open-source \name{}~\cite{swics}, the first  virtual and modular security testbed which interconnects industrial components over a realistically simulated 5G network, enabling reproducible research and facilitating future studies. 
    \item By replicating well-known network-level attacks, we compare the resilience of 5G-enabled and wired \acp{ICS}, showing that the robustness of 5G depends on channel conditions~(\S\ref{sec:impactPhy}).
    \item By deploying and evaluating two communication-based intrusion detection systems, well-established security controls often used in \acp{ICS}, we show that traditional security approaches previously considered sufficient may no longer be effective~(\S\ref{sec:ids}). 
    \item We explore the wireless interface  to study passive reconnaissance and jamming attacks, highlighting that disruptions mainly threaten availability\textemdash{}the most critical security aspect of \acp{ICS}\textemdash{}to reveal the limitations of 5G in \ac{ICS}~(\S\ref{sec:wirelessAttacks}).
\end{enumerate}

\noindent\textbf{Availability:}
Source code and artifacts are available at
\url{github.com/RWTH-SPICe/SWICS} and 
\url{doi.org/10.5281/zenodo.19550997}

\section{Security of (Wireless) ICSs}

To lay the foundation for our work, we introduce the core concepts of \acp{ICS}~(\S\ref{sec:background:ics}) and explain how 5G, with its ability to operate in the mmWave spectrum, enables sub-1~ms latencies~(\S\ref{sec:background:5g}). 
We then review prior research on the security implications of 5G in \acp{ICS} and show that existing testbeds do not support comprehensive studies of integrating 5G into \acp{ICS}, highlighting our contributions~(\S\ref{sec:related}).

\subsection{Industrial Control Systems}
\label{sec:background:ics}

\acfp{ICS} form the backbone of modern production and critical infrastructure by linking physical components with digital control and monitoring systems~\cite{knapp2024_CH3}.
They are structured into three levels: \emph{fieldbus}, \emph{control}, and \emph{supervisory}.
At the fieldbus level, sensors monitor the physical environment and relay process values to the control level, while actuators such as motors or valves receive low-level control signals to directly influence the physical process.
Controllers, e.g., \acp{PLC}, interpret sensor data and compute the commands that drive these actuators, forming together a continuous sensing–actuation loop known as the process cycle.
Controllers also report the current process state to the supervisory level, e.g., a \ac{HMI}, enabling humans to oversee and steer operations.

Because of their tight coupling with the physical world, \acp{ICS} must meet stringent safety and availability requirements~\cite{michaelides2025industry5G}, demanding strict network performance guarantees, especially low latency, minimal jitter, and low packet loss~\cite{michaelides2025industry5G}.
Due to the increasing connectivity of \acp{ICS} to the Internet~\cite{knapp2024_CH3}, it is more important than ever to develop security measures that do not interfere with these requirements.
Thus, to study ICS security, research relies on \emph{testbeds}, i.e., physical or virtual replicas of real systems, to avoid harming actual infrastructure~\cite{conti2021survey}.
Unlike physical testbeds, virtual testbeds eliminate the need for expensive hardware, enhance reproducibility, and allow for controlled experiments.

\subsection{Wireless Communication in ICSs}
\label{sec:background:5g}

Driven by demands for increased automation, flexibility, and efficiency in production and critical infrastructure, \acp{ICS} are currently shifting from traditional wired to wireless communication~\cite{michaelides2025industry5G}.
While several wireless technologies are suitable for industrial use cases, the fifth generation of cellular networks (5G)  offers sub-millisecond latencies, making it particularly promising for industrial applications with strict latency requirements~\cite{michaelides2025industry5G, 5gacia_use_cases}.

In contrast to most wireless technologies, 5G comprises three main components: \acp{UE}, \acp{gNB}, and the Core Network (CN). \acp{UE}, such as sensors, \ac{PLC}s, or \acp{HMI}, act as endpoints that generate and consume data, while \acp{gNB} provide the radio access interface and connect \acp{UE} to the CN, which handles data routing and connection management. 
Notably, 5G distinguishes between control and user plane traffic: control plane messages, which manage signaling and system operation, are mandatorily integrity protected, whereas user plane traffic (i.e., the industrial data) is not~\cite{michaelides2025industry5G}. 
This distinction has important security implications, as enabling integrity protection for user plane traffic introduces additional overhead that may conflict with stringent sub-millisecond latency requirements~\cite{michaelides2026ransecurity}.
For our analysis of the \emph{wireless} channel in \acp{ICS}, we abstract from this architectural complexity and focus on \emph{\acp{UE}}, \emph{\acp{gNB}}, and the \emph{wireless physical channel} between them, where these trade-offs become most relevant.

A key advancement of 5G is its support for mmWaves (>\SI{24}{GHz}), which enable significantly shorter transmission intervals, as low as \SI{15.625}{\micro\second}, compared to sub-\SI{6}{GHz} bands with slot durations around \SI{0.25}{ms}~\cite{mmWavegood}. 
This finer scheduling granularity enables end-to-end latencies below \SI{1}{\milli\second}, making 5G well-suited for latency-critical \acp{ICS}, particularly for applications requiring tight timing and high reliability, in contrast to alternatives such as Wi-Fi~\cite{michaelides2025industry5G}.

\noindent\textbf{Challenges in mmWave Environments.} While mmWave is essential for achieving sub-ms latency, it introduces challenges such as high propagation loss, limited coverage, strong sensitivity to blockage, and susceptibility to environmental factors such as the weather~\cite{abuhdima2021impactweatherconditions5g}. 
Due to these limitations, line-of-sight (LoS) conditions between \acp{UE} and \acp{gNB} are often necessary. 
These issues are mitigated through dense cell deployments that reduce transmission distances, as well as techniques such as MIMO and beamforming, which improve reliability and capacity by using multiple antennas and directing signals efficiently~\cite{birutis2022study5G}. 
Although less reliable, non-line-of-sight (NLoS) communication remains possible via signal reflections~\cite{9318411}. 
In industrial settings, these challenges are more manageable due to controlled environments, particularly in stationary \acp{ICS} that enable stable LoS links and precise beamforming.

\subsection{Related Work}
\label{sec:related}

Although the security of \ac{ICS} has received significant research attention in recent years (e.g.,~\cite{rodofile, watertank, MOHAMMED2023103007, 10905487,10.1145/3140241.3140253,299553}), the implications of transitioning from wired to 5G communication remain underexplored.
Previous work~\cite{michaelides2025industry5G,5gacia3} discusses the integration of 5G  into \acp{ICS} and the associated security challenges, but only on a theoretical level. 
We attribute this gap to the lack of testbeds that integrate 5G (mmWave) communication in \acp{ICS}.
In the following, we review existing wireless ICS testbeds to highlight this deficiency, then analyze studies on ICS security in traditional wired settings, and finally discuss research on 5G security in general-purpose applications.

\noindent
\textbf{Testbeds for ICS Security.}
Testbeds are crucial for safely evaluating cyber-physical attacks within \ac{ICS} environments (cf.~§\ref{sec:background:ics}).
Although numerous ICS security testbeds exist, they predominantly rely on wired communication or focus on wireless protocols, which are less promising than 5G mmWave.
For example, Conti et al.~\cite{conti2021survey} surveyed 61 ICS security testbeds, all of which use only wired communication.
While Kampourakis et al.~\cite{kampourakis2023survey} examined 46 wireless testbeds in cyber-physical systems, only four specifically target industrial processes, and none employs cellular communication (4G/5G).
Similarly, in the IoT context, De Santana et al.~\cite{deSantana2024cybersecurity} identified only three industrial testbeds involving 4G/5G technology.

Table~\ref{tab:ics_wireless_testbeds} summarizes these existing wireless testbeds relevant to ICS security. 
As shown, no existing testbed relies on 5G mmWave for communication~\cite{prakash2017Fingerprinting, adepu2017SWATJamming, tomic2018Waterbox,yasai2020IotCad,gardiner2019SecurityTestbeds,oliver2018testbed5G,lee2021factory}. 
Thus, experiments investigating the wireless channel may not be transferable to 5G mmWave due to its sub-\SI{1}{ms} latency capabilities.
Lastly, none of the existing testbeds allow reproducible research, e.g., evaluating the impact of different attacks under identical wireless channel conditions and physical process state.
To bridge this gap, we introduce \name{}, the first virtual testbed that integrates 5G mmWave into an \ac{ICS} and fully relies on simulation to enable replicable research.

\noindent
\textbf{Security Experiments on ICS.}
\label{sec:5g-security}
Traditional \ac{ICS} research has extensively studied the impact of cyberattacks on physical processes.
Rodofile et al.~\cite{rodofile} conducted injection and flooding attacks on a mining refinery.
Munteanu et al.~\cite{watertank} analyzed attacks on sensors and actuators in a water tank system, using statistical model checking to evaluate their impact and intrusion detection performance.
Mohammed et al.~\cite{MOHAMMED2023103007} examined \ac{DOS} attacks on oil and gas infrastructure, assessing impact and detection methods.
These are among many other works such as~\cite{10905487,10.1145/3140241.3140253,299553}, exploring attacks on \acp{ICS}.
These studies further highlight the lack of focus on the security implications of transitioning to 5G-enabled \acp{ICS}.

\noindent
\textbf{Studies on 5G Security.}
While related work does not specifically assess the practical security implications of transitioning to 5G for \acp{ICS}, different research streams have studied 5G security in general-purpose settings.
For example, Ludant et al.~\cite{10.1145/3448300.3467817} demonstrated low-power DoS attacks on 5G's physical layer under a strong attacker model, and Birutis et al.~\cite{jammingnNorway} investigated jamming in commercial 5G, measuring bandwidth degradation but not latency, which is critical in industrial settings.
Previous generations, such as LTE~\cite{7069323,sigover}, have also been studied without considering ICS impact.
Nevertheless, these works highlight the need to consider wireless-medium vulnerabilities when analyzing \acp{ICS} security and support the transition to 5G for evolving industrial demands~\cite{michaelides2025industry5G}.
Lastly, studies exploring attack vectors introduced by other wireless technologies in \ac{ICS}, e.g., Wi-Fi~\cite{adepu2017SWATJamming, tomic2018Waterbox, yasai2020IotCad, prakash2017Fingerprinting} may not be transferable to mmWave.

\begin{takeaway}
    Despite the transition to wireless \acp{ICS}, research on the security implications remains limited, largely due to the lack of publicly available testbeds, ultimately highlighting the need for a virtual 5G-\ac{ICS} testbed, that enables reproducible research.
\end{takeaway}

\begin{table}[t]
    \centering
    \footnotesize
    \caption{The current landscape of security testbeds for wireless ICS lacks a platform that supports 5G mmWave.}
    \label{tab:ics_wireless_testbeds}
        \begin{tabular}{@{} l c c c c @{}}
        \toprule
        \textbf{Study} & \textbf{ICS Focus} & \textbf{Tech.} & \textbf{Real Process} & \textbf{Reproducibility} \\
        \midrule
        Prakash et al.~\cite{prakash2017Fingerprinting} & Yes & Wi-Fi\textsuperscript{$\delta$} & Yes & No \\
        Adepu et al.~\cite{adepu2017SWATJamming} & Yes & Wi-Fi\textsuperscript{$\delta$} & Yes & No \\
        Tomić et al.~\cite{tomic2018Waterbox} & Yes & Wi-Fi\textsuperscript{$\delta$} & Yes & No \\
        Yasaei et al.~\cite{yasai2020IotCad} & Yes & Wi-Fi\textsuperscript{$\delta$} & Yes\textsuperscript{$\alpha$} & No \\
        Gardiner et al.~\cite{gardiner2019SecurityTestbeds} & Yes & 4G\textsuperscript{$\delta$} & Yes\textsuperscript{$\beta$} & No \\
        Oliver et al.~\cite{oliver2018testbed5G} & No & 5G\textsuperscript{$\delta$} & Yes & No \\
        Lee et al.~\cite{lee2021factory} & Yes & 5G\textsuperscript{$\delta$} & No\textsuperscript{$\gamma$} & No \\
        \midrule
        \textit{\name{} (this work)} & Yes & 5G & Yes\textsuperscript{$\alpha$} & Yes \\
        \bottomrule
        \end{tabular}%
    \vspace{0.5em}
    \raggedright
    \textsuperscript{$\alpha$}Simulated\quad \quad \quad \quad 
    \textsuperscript{$\beta$}Subprocess\quad \quad \quad \quad 
    \textsuperscript{$\gamma$}Replay traffic \quad \quad \quad \quad 
    \textsuperscript{$\delta$}sub-\SI{6}{GHz}
\end{table}

\section{\name{} -- Simulating Wireless Communication in Industrial Control Systems}
\label{sec:process}

To address the lack of testbeds that combine industrial physical processes with 5G, and to enable reproducible research, we introduce \name{}, a fully virtual, simulation-based ICS testbed.
In this section, we motivate the use of simulation and discuss its benefits~(\S\ref{sec:virtual}).
We then present the design of the simulated \ac{ICS}, an enhanced version of a state-of-the-art industrial testbed, and describe its integration into the ns-3 simulator~(\S\ref{sec:swics}). 
Finally, we detail our realistic configuration of the 5G network  reflecting real-world deployments~(\S\ref{sec:5g_imp}).

\subsection{The Benefits of Simulation-based Testbeds}
\label{sec:virtual}
Comprehensively analyzing the security of wireless industrial systems presents several major challenges.
First, deploying attacks or testing defense mechanisms on real industrial systems is inherently risky, as such actions may disrupt production, critical processes, or even endanger human safety.
Second, introducing a wireless communication channel further complicates experimentation:
Wireless conditions fluctuate due to seemingly insignificant and uncontrollable factors such as dust particles (cf. §\ref{sec:background:5g}).
These fluctuations make it nearly impossible to recreate identical radio environments across experiments.
More importantly, the inherent non-determinism of the physical wireless medium prevents observed effects from being reliably attributed to specific system changes, attacks, or defenses.
This lack of traceability and consistency undermines fair evaluation, reproducibility, and confidence in experimental results.
While constructing small-scale physical testbeds is possible, it is extremely costly and still offers limited control over environmental conditions.

On the other hand, a fully virtual testbed enables safe and replicable experimentation with precise control over industrial networks.
Consequently, as the foundation for our research on the security implications of wireless communication in industrial systems, we develop \name{}~\cite{swics}, the first virtual testbed to interconnect \ac{ICS} components over 5G.
\name{} follows the discrete-event simulation (DES) paradigm to accurately model an \ac{ICS}, where industrial devices control a responsive physical process communicating via 5G.
Leveraging DES enables fine-grained, deterministic control over both physical and wireless dynamics, allowing us to isolate factors such as communication type or attack presence and reliably attribute any differing outcomes to isolated system parameters.

\subsection{Simulation of the Industrial Control System}
\label{sec:swics}

\begin{figure}[t]
    \centering
    \includegraphics[width=0.8\columnwidth]{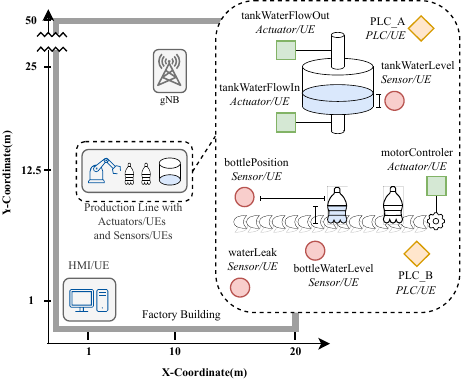}
    \caption{
        \name{} simulates a bottle-filling plant~\cite{dehlaghi2023icssim}, comprising a water tank, valve-actuators, fill-level and spill sensors, a conveyor belt, an HMI, two PLCs, and 5G communication.
    }
    \label{fig:process}
\end{figure}

At the core of \name{} sits the physical process of the simulated \ac{ICS}.
To enable the rapid execution of multiple attacks, we require a physical process that is inherently robust and can quickly recover from attacks.
Consequently, we adapt the existing design of a simulated water bottle-filling plant~\cite{dietz2020integrating} relying on a conveyor belt for automated bottle transport~\cite{dehlaghi2023icssim}. 
We further advance this design by increasing both the physical realism and operational complexity of the system.
Fig.~\ref{fig:process} visualizes our resulting 5G-enabled \ac{ICS}.

\noindent\textbf{ICS Operation and Enhancements.} The \ac{ICS} controls two central functions of the bottle-filling process: positioning bottles beneath the water tank using a conveyor belt and regulating water inflow to the tank and outflow into bottles (cf.\ Fig.~\ref{fig:process}). 
Two \acp{PLC} and one HMI coordinate these tasks. 
PLC\_A regulates water flow by opening the output valve when a bottle is detected and closing it once the desired fill level is reached. 
It also monitors the tank level sensor to refill the tank as needed by opening the input valve.
PLC\_B manages conveyor movement based on the sensors for bottle position and fill level, stopping the belt when an empty bottle arrives beneath the tank and restarting it once filling is complete. 
The HMI collects an overview of the process state from PLC\_A and PLC\_B.

To improve realism beyond prior work, we extend the simulated process in four ways. 
First, we add a water-leak sensor that allows PLC\_A to detect abnormal outflow conditions. 
Second, we introduce measurement noise by applying up to 0.5\% random error to all sensor readings, reflecting imperfections common in industrial environments. 
Third, we replace fixed liquid flow-rate approximations with a physically accurate model based on Torricelli’s law~\cite{tec-science:Torricelli}, making fill times dependent on the current tank level.
Finally, we integrate an HMI that periodically polls the system state and logs timeouts, enabling operator visibility.
Lastly, to ensure safe operation, we additionally implement an emergency mechanism: if sensor readings cease for a predefined interval, the PLCs halt the process by closing all valves and stopping the conveyor belt.
Together, these enhancements yield a more realistic, fault-aware, and operationally complete representation of an industrial bottle-filling system.

\noindent
\textbf{Implementation.} 
We implement our \ac{ICS} in the \emph{ns-3} simulator, with all components realized as \emph{ns-3 Applications}. 
This approach allows seamless integration with existing ns-3 classes, including those for TCP/IP communication and the deployment of applications on any \emph{ns-3 Node}. 
It also enables flexible and rapid addition of new components, making \name{} an extendable testbed for future experiments across adifferent scenarios.  We also extend \emph{ns-3} to fully support the Modbus protocol, which is widely used in \acp{ICS}~\cite{conti2021survey}. %

\begin{figure}[t]
    \centering
    \begin{minipage}{0.45\textwidth}
        \centering
        \includegraphics[width=\linewidth]{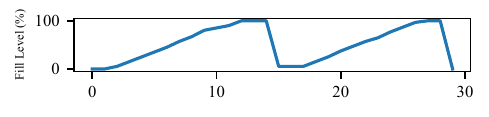}
        \\[-2ex]
        \footnotesize (a) Reference Physical Process 
    \end{minipage}
    \hfill
    \begin{minipage}{0.45\textwidth}
        \centering
        \includegraphics[width=\linewidth]{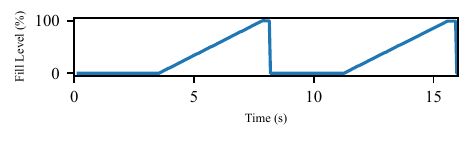}
        \\[-2ex]
        \footnotesize (b) Our Physical Process
    \end{minipage}%
    \caption{\name{}’s behavior (b) closely matches the original reference simulated process~\cite{dietz2020integrating} (a). To improve realism, we smooth bottle filling by increasing the polling rate and model liquid flow using Torricelli’s law, yielding more realistic filling times and reflecting the natural acceleration of liquid.}
    \label{fig:sensors}
\end{figure}

\noindent
\textbf{Accuracy.}
To ensure accuracy of our physical process, we analyze its behavior over time. 
Fig. \ref{fig:sensors} (b) exhibits the expected behavior: as bottles fill with liquid, the fill level sensor detects an increasing level, and the corresponding measurement level rise. 
Once a bottle is filled, it moves forward to accommodate a new empty bottle, causing the sensor to report an empty state. 
The time interval between consecutive bottles is reflected in the flat portions of the graph. 
We validate our implementation by comparing it to the original process (cf. Fig.~\ref{fig:sensors} (a)), observing comparable overall behavior, while \name{} shows smoother sensor readings due to a higher polling rate.

\subsection{Integration of 5G Communication}
\label{sec:5g_imp}
By simulating a physical process and leveraging the deterministic nature of DES, \name{} provides a solid foundation for studying the security implications of 5G in \acp{ICS}.
To simulate 5G, we use a state-of-the-art ns-3 module and configure it in accordance with relevant studies to ensure high accuracy.
\sm{check above}

\noindent
\textbf{mmWave Module.} We add 5G communication between the \ac{ICS} components by utilizing the \emph{mmWave} ns-3 3GPPP-compliant extension~\cite{mmwave}, which is widely used in academia (e.g.,~\cite{patriciello2019E2e5g, raca2020dataset, lacava2024oran}).
This extension implements 5G-specific behavior at the PHY and MAC layers on top of the LTE module to enable end-to-end simulation of 5G mmWave networks.
As such, it enhances LTE eNBs and UEs with mmWave capabilities by supporting, e.g., 5G frame structure, transmission/reception, and beamforming, allowing realistic simulation of communication in mmWave spectrum.
While ns-3 currently lacks a full-stack 5G implementation, the mmWave module fully meets our research needs, as we focus solely on studying the impact of the 5G channel in \acp{ICS}.

\noindent
\textbf{Topology.}
Our network consists of 10 UEs, one for each industrial device (sensors, actuators, PLCs, HMI), positioned in a factory-sized building measuring \SI{50}{m} $\times$ \SI{20}{m} $\times$ \SI{10}{m} (cf.\ Fig.~\ref{fig:process}).
Further, we place one gNB within the facility that is connected to the core network, facilitating communication between the devices.
All UEs are stationary at ground level, while the gNB has a height of 9.5 m, enabling LoS between them. 
The gNB utilizes an 8$\times$8 MIMO configuration and the UEs a 2$\times$2 MIMO configuration, both employing beamforming to overcome challenges w.r.t. mmWaves~(cf.\S\ref{sec:background:5g}).

\noindent
\textbf{Realism.} To ensure accurate and realistic network behavior, we configure our network based on 3GPP specification and prior works.
First, we set our network to operate on band n257 (\SI{28}{GHz}), which is well suited for industrial applications due to its low latency~\cite{gsma2022}.
We configure the gNB transmission power to \SI{34}{dBm}, in line with 
3GPP specifications, which define transmission power for medium-range base stations up to \SI{38}{dBm}~\cite[§6.2]{3gpp38104}.
For UEs, band n257 encompasses all power categories (1-7) defined by 3GPP~\cite[§6.2.1]{3GPPTS38521}.
We select power category 3 (\SI{23}{dBm}), designed for applications requiring lower power output, such as industrial devices.
To account for effects such as signal attenuation in industrial environments, we use the 3GPP indoor factory channel model~\cite[§7]{3GPPTR3801,10.1145/3592149.3592155}.

\begin{takeaway} 
    To investigate the impact of 5G on \ac{ICS} security, we introduce \name{} which accurately simulates both the physical process and wireless 5G channel.
    Thus, we ensure that our experiments closely reflect real-world results.
\end{takeaway}

\section{Analysis Methodology}
\label{sec:methodology}

To analyze the implications of transitioning from wired to 5G communication in \ac{ICS}, we deploy \name{} in both a 5G mmWave configuration and a traditional Ethernet setup. (cf.\ §\ref{sec:process}).
Then, we compare the functionality of the \ac{ICS} utilizing both media under benign and attack conditions. %
By maintaining consistent parameterization across all deployments, we ensure that any observed differences are solely attributable to the communication medium.

In addition to the physical medium, we evaluate changing channel conditions, which are a critical factor influencing network performance, especially considering that 5G mmWave deployments operate at very high frequencies rendering them particularly sensitive to environmental impact (cf.~§\ref{sec:background:5g}).
Consequently, in our experiments, we consider three deployment scenarios:

\emph{\textbf{Wired.}} A wired version of the deployment described in §\ref{sec:process} where all 5G connections are replaced with 100\,Mbit/s Ethernet, serving as baseline for comparison.

\emph{\textbf{ \goodChannel{}.}} The 5G setup as described in §\ref{sec:process} operating under optimal channel conditions.

\emph{\textbf{\badChannel{}.}} The same 5G setup as before with intentionally degraded channel conditions, achieved by increasing noise to emulate interference from a device operating on the same area and frequency.

Using these scenarios, we evaluate how the communication medium influences the impact of network-level attacks.
To this end, we first define a threat model based on realistic assumptions and prior work (\S\ref{sec:threat}), ensuring it is robust and applicable to real-world 5G \acp{ICS}. We then create a baseline scenario in which \name{}’s physical process operates without interference (\S\ref{sec:baseline}). Building on this benign scenario, we execute well-known attacks\textemdash{}common in publicly available studies and datasets\textemdash{}against the \acp{ICS} (\S\ref{sec:attacks}). To assess the security implications of each deployment, we compare differences in each attack regarding their direct effect on the physical process (§\ref{sec:impactPhy}), the performance of existing security measures (§\ref{sec:ids}), and novel attack vectors of the wireless channel (§\ref{sec:wirelessAttacks}).

\begin{figure}
    \centering
    \begin{minipage}{0.45\textwidth}
        \centering
        \includegraphics[height=2.75cm,width=\linewidth]{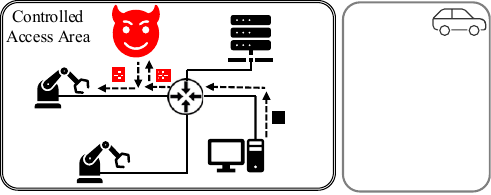}
        \footnotesize (a) Threat Model in Wired Setup
    \end{minipage}%
    \hfill
    \\[1ex]
    \begin{minipage}{0.45\textwidth}
        \centering
        \includegraphics[height=3.25cm,width=\linewidth]{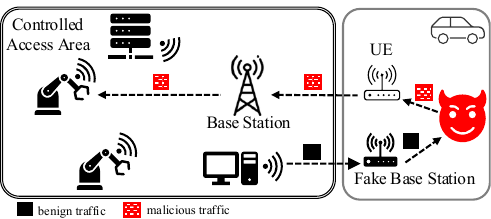}
        \footnotesize (b) Threat Model in 5G Setup
    \end{minipage}
\caption{
    We assume that the attacker can modify ICS traffic the \emph{wired} (a) and \emph{5G} (b) deployments, although the means to achieve these capabilities depend on the physical medium.
}
\label{fig:threat_model}
\end{figure}

\subsection{Threat Model and Scope}
\label{sec:threat}
Before assessing 5G's impact on \ac{ICS} security, we define our threat model consideing two different attacks with distinct capabilities:

\noindent\textbf{Insider.}
To highlight the potential impact of insufficient security in \acp{ICS}, we consider an attacker capable of injecting, blocking, and modifying packets in both deployments. 
Achieving these capabilities requires access to the \ac{ICS} network, depending on the underlying communication medium (Fig.~\ref{fig:threat_model}).
As \acp{ICS} prioritize operational goals, security controls are often omitted with recent studies showing that only 6.5\% of industrial devices use TLS with  ~42\% being misconfigured~\cite{IOTTLS}. 
Therefore, we assume the absence of \emph{end-to-end security} in wired industrial deployments.
Thus, in such settings (Fig.~\ref{fig:threat_model}~(a)), physical or logical access to a network interface—e.g., via a compromised port, a completed ICS cyber kill chain~\cite{ICSkillChain}, or a misconfigured TLS client—is sufficient to obtain these capabilities.
We further assume that such capabilities remain achievable in 5G-enabled ICS, as it remains unclear if adoption of end-to-end security will increase when switching to 5G.
Moreover, 5G user plane security controls may be disabled in favor of sub-\SI{1}{ms} communication requirements~\cite{michaelides2026ransecurity}.
Even when user plane security is enabled, attackers may exploit vulnerable ICS endpoints~\cite{wang2023plcSurvey} or unsecured wired segments in hybrid deployments routed through the 5G network.
In addition, the 5G infrastructure introduces further attack vectors: an adversary may exploit the open wireless interface—e.g., via signal overshadowing~\cite{sigover}—to deploy a rogue base station (Fig.~\ref{fig:threat_model}(b)), which can then be used to relay, intercept, and modify traffic, effectively acting as a machine-in-the-middle~\cite{brekainglte}.

\noindent\textbf{Outsider.}
To further illustrate the security implications of 5G for \acp{ICS}, we consider a threat model for wireless deployments under the assumption of a \textit{perfectly secured ICS}. 
In this setting, the attacker is limited to interacting with the (encrypted) radio channel over the open air interface, with no access to internal system components or interfaces. 
We use this model to highlight security complications introduced by wireless communication in §\ref{sec:wirelessAttacks}.
    
Ultimately, this threat model emphasizes the need for security-by-design in future 5G-enabled ICS, as also highlighted by the European Union Agency for Cybersecurity~\cite{enisa}.
We further elaborate on the implications of our assumptions in §\ref{sec:discussion}.

\subsection{Benign Scenario}
\label{sec:baseline}

To serve as a reference for our experiments, we establish an attack-free scenario which solely contains benign process behavior with a total duration of 20 minutes.
In this scenario, bottles are being filled normally for 10 minutes.
After that point, to introduce variability, the HMI halts the physical process, causing the conveyor belt to stop. 
After 1 minute, the HMI issues a command to restart the process.
Upon restarting, the physical process resumes in two stages, initially operating at half speed before gradually returning to normal speed after another 2 minutes. 
These fluctuations introduce complexity by disrupting the otherwise deterministic traffic patterns, thereby creating a more realistic reference scenario and also aligning with the diversity requirement of datasets for security evaluation~\cite{conti2021survey}.

\subsection{Attack Scenarios}
\label{sec:attacks}

Following Conti et al.~\cite{conti2021survey}, we implement four representative cyber-attack scenarios an \emph{insider} can perform: \ac{DOS}, Machine-in-the-Middle, Suppression, and Injection . 
While well studied in wired \acp{ICS} (cf.~§\ref{sec:related}), their impact on the physical process in 5G remains largely unexplored, particularly how the communication medium affects attack behavior and consequences.
To investigate this, we use \name{} to compare process stability under both wired and 5G attack scenarios. For each attack, we schedule multiple variants with increasing durations on the baseline scenario. Leveraging the deterministic nature of discrete event simulation, attacks are precisely scheduled at \SI{200}{s}, \SI{400}{s}, \SI{600}{s}, \SI{800}{s}, and \SI{1000}{s}. This ensures identical process states and enables a clear comparison across scenarios.

\noindent
\textbf{Denial of Service (DoS).}
Flooding attacks are among the most common types of \ac{DOS} attacks.
They typically involve sending a large volume of data to a target, exhausting the network’s capacity or overwhelming its computational resources, rendering it unable to process legitimate traffic.
This attack type is easy to set up and poses a significant threat to industrial processes, indiscriminately disrupting normal operations without requiring extensive knowledge of the process, leading to unpredictable behavior.
We implement the \ac{DOS} attack by simulating an attacker retransmitting numerous Modbus read requests from the \ac{HMI} to PLC\_A during the given attack schedules.
We intentionally set the number of injected packets so that the attack affects the physical process without causing a water spill to remain as subtle as possible.

\noindent
\textbf{Machine in the Middle (MitM).} 
In this attack, the attacker exploits the absence of integrity protection\textemdash{}common in \acp{ICS}~\cite{henze2017network}\textemdash{}to modify legitimate packets.
By positioning themselves on the communication path, the attacker can intercept and alter packets such as sensor readings or PLC commands.
While such attacks are straightforward in wired networks (e.g., via a physical MiTM or ARP spoofing), they have also been shown to be feasible in cellular communication if integrity protection is absent~\cite{brekainglte}.
To implement this attack, we send altered commands to the \textit{motionController} regulating the conveyor belt by replacing  \emph{stop commands} with a \emph{keep moving} command.
Thus, we prevent the conveyor belt from stopping and potentially cause liquid spillage as bottles do not stop beneath the filling station.

\noindent
\textbf{Injection.}
To conduct an injection attack, an attacker only requires access to the communication network.
Exploiting the lack of authentication, the attacker can then inject messages into the network, pretending to be one of the two communicating parties, potentially sending unexpected commands or false sensor readings. 
If the channel provides origin authentication, the attacker can still intercept legitimate packets and re-inject them to the destination at a later time (a.k.a. a replay attack).
In our scenarios, we implement the replay variant by re‑transmitting \emph{open} commands to the tank’s liquid output valve from PLC\_A. Depending on the physical process state\textemdash{}i.e., whether a bottle is positioned under the valve\textemdash{}the attack may cause liquid spills.

\noindent
\textbf{Suppression.} In contrast to previous attacks, where an attacker manipulates legitimate communication to compromise the system, a suppression attack disrupts the physical process by selectively dropping packets, such as PLC commands. 
This can be achieved by compromising either the sender/receiver, infiltrating the network somewhere along the path between them, or in 5G through methods such as  smart reactive jamming~\cite{jamming}.
We simulate this attack by blocking the PLC’s network interface during specified time frames, preventing it from transmitting any messages. As a result, commands such as opening/closing valves are not issued during these intervals, potentially causing spills.

\begin{takeaway}
    Using \name{} we study the impact of 5G on ICS security by comparing the effects of well-established attacks. By using consistent parameterization, we ensure that any diverging behavior can be attributed solely to the physical medium.
\end{takeaway}

\section{Impact on the Physical Process}
\label{sec:impactPhy}

\begin{figure*}
    \centering
    \includegraphics[width=\textwidth]{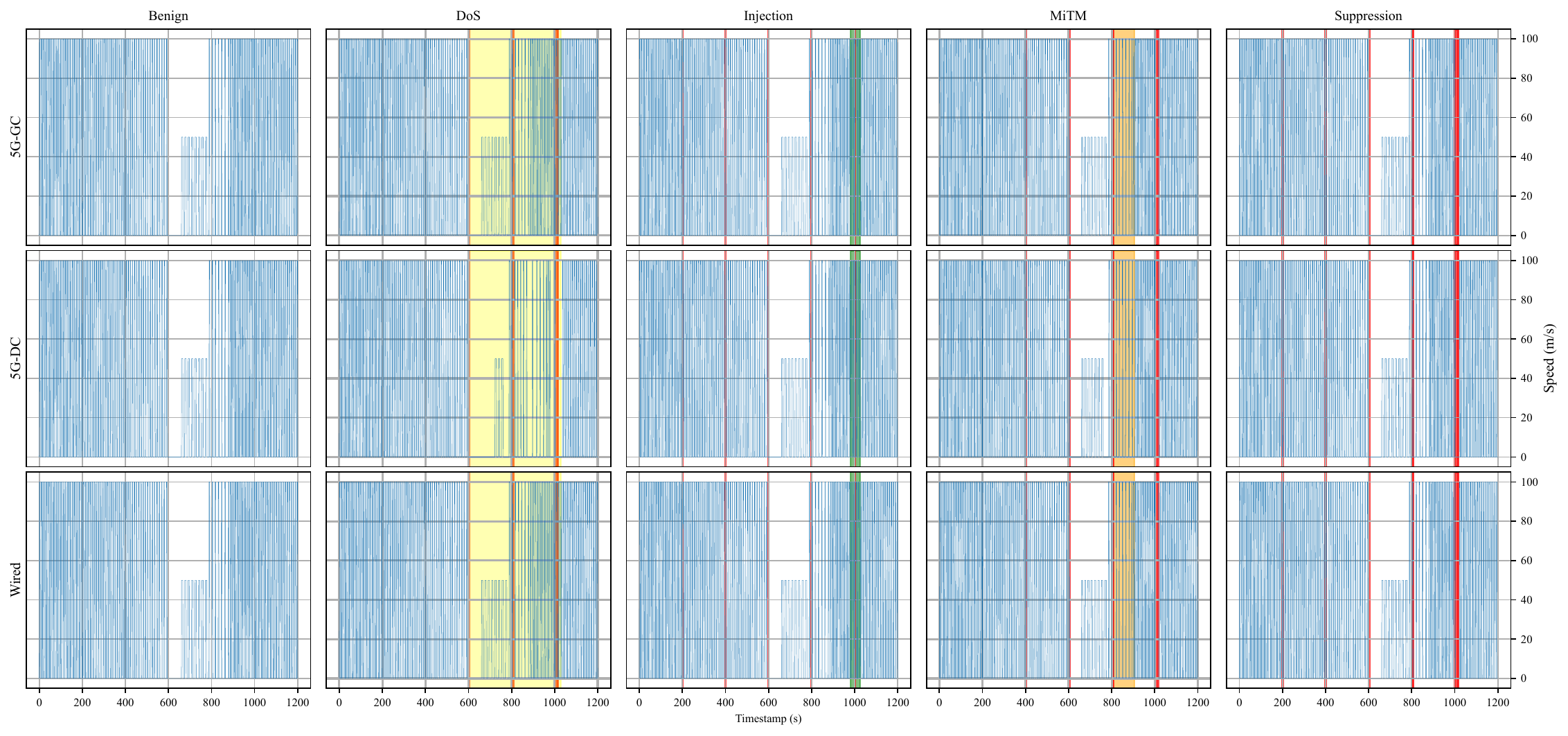}
    \caption{The conveyor belt speed in the \emph{Wired}, \emph{\goodChannel{}}, and \emph{\badChannel{}} deployments is stable under benign conditions.
    Under attack, \emph{Wired} and \emph{\goodChannel{}} show similar resilience with only minor jitter-induced deviations, whereas the noisy \emph{\badChannel{}} deployment significantly amplifies the effects of DoS and MiTM attacks due to increased jitter, packet loss, and latency.}
    \label{fig:Attacks_comparison}
\end{figure*}

The primary goal of \ac{ICS} security is to prevent and minimize damage to the physical process. 
After outlining our methodology, we evaluate the impact of 5G on this process by deploying each attack described previously across all scenarios.
We then evaluate differences in the behavior of the physical process (i.e., conveyor belt speed or water spills) compared to its expected behavior. By monitoring these attributes, we can quantify the effects of attacks, utilizing the number of liquid spills as a measure of critical failure.
In this section, we review all attacks and report differences in their effects across the three scenarios. Due to \name{}'s determinism and consistent parameterization, we can attribute any observed differences solely to the communication medium.

\noindent
\textbf{Results.} To show the impact on the physical process, Fig.\ref{fig:Attacks_comparison} visualizes the conveyor belt speed over time for each scenario. 
The {\colorbox{softred}{red}}-shaded regions in the plots denote active attack periods, allowing for a clear comparison of the process response during and after disruptions. 
Further, Tab.\,\ref{tab:water_spills} complements this by summarizing the amount of liquid spilled as detected by the corresponding sensor.

\noindent
\textbf{Benign.} 
In this configuration, the physical process runs smoothly and identically across all scenarios.
Even for \badChannel{}, the system's robustness and high sensor polling rate ensure that packet losses and increased jitter caused by noise do not affect process operation.

\begin{table}[t]
    \centering
    \footnotesize
    \caption{
        Liquid spills as a measure of critical failure show the increased effect of attacks against the \badChannel{} scenario.} %
    \label{tab:water_spills}
    \begin{tabularx}{\columnwidth}{l *{5}{>{\centering\arraybackslash}X}}
    \toprule
     & Benign & DoS & Injection & MitM & Suppr. \\
    \midrule
    Wired    & 0      & 0                & 0          & 4    & 2           \\
    \goodChannel{}       & 0      & 0                & 1          & 4    & 2           \\
    \badChannel{}    & 0      & 1 (long)  & 0          & 5    & 2           \\
    \bottomrule
    \end{tabularx}
\end{table}

\noindent
\textbf{DoS.}
Similarly to the benign scenario, during the DoS attack, we observe no substantial differences between \goodChannel{} and Wired.
All short attack variants appear to have no impact on the physical process, likely due to its inherent robustness.
However, when considering the \badChannel{}, particularly from the third attack variant onward ({\colorbox{myYellow}{yellow}} in Fig.~\ref{fig:Attacks_comparison}), the attack clearly impacts the physical process. 
The third attack variant coincides with the moment the PLC issues a command to stop the conveyor belt, which continues to move briefly afterward at half speed.
Although the stop command from the PLC reaches the conveyor belt and it eventually stops, the flooding attack causes queues to overflow, resulting in excessive packet drops and triggering retransmissions. 
This effect, combined with the noisy 5G channel that introduces additional packet losses, leads to delays in the delivery of subsequent PLC commands, e.g., to restart the conveyor.
The increased latency in PLC start/stop commands becomes more evident in the fourth attack variant, where the conveyor belt experiences extended and inconsistent periods of motion and halting.
Finally, for the fifth, most intense attack variant, communication is completely disrupted, resulting in a total loss of availability.
In the \badChannel{}, this variant also leads to a long liquid spill, as delayed control signals prevent timely process control.

\noindent
\textbf{Injection.}
The injection attack is designed to open the liquid valve at random time points, potentially causing spills. 
Our results reveal an unexpected outcome.
In the Wired setup, the attack has no effect because the injected packet arrives immediately before a legitimate update, which nullifies the attack's effect before it is registered by the system. 
In the \goodChannel{} scenario, however, the legitimate packet arrives later due to increased delays, causing a water spill.
In the \badChannel{} deployment, an earlier burst of packet loss causes a shift in the update cycle, which coincidentally results in the legitimate update packet to arrive just before system reading; effectively nullifying the attack similar to the Wired deployment by pure chance.

\noindent
\textbf{MitM.}
The MiTM attack targets the conveyor belt PLC, aiming at disrupting timing-critical operations such as placing bottles under the valve.
As shown in Tab.\,~\ref{tab:water_spills}, both Wired and \goodChannel{} setups experience four spills, while 5G-DC sees one additional spill—all occurring during attacks.
The difference lies in how each network handles the third attack variant. 
In Wired and \goodChannel{}, the attack command arrives just before a legitimate stop command, which quickly overrides it due to the controller’s polling rate—limiting the impact. 
In \badChannel{}, however, higher jitter and MiTM processing delays cause the altered packet to arrive too late, making the spill unavoidable.
The noisy channel also delays benign PLC instructions causing the conveyor belt to remain at a lower speed for extra process cycles compared to the other scenarios ({\colorbox{myOrange}{orange}} in Fig.~\ref{fig:Attacks_comparison}).

\noindent
\textbf{Suppression.}
During the suppression attack, we observe identical effects across all scenarios. 
Because the attacker drops all packets, additional latency and occasional packet loss caused by network conditions have no substantial impact. 
As a result, each scenario experiences two liquid spills. 
In the final variant, the attack escalates to a DoS, triggering the \name{}'s safety mechanism ultimately halting the production process~(cf.~§\ref{sec:swics}).

\begin{takeaway} 
    While reliable 5G channels can match the resilience of wired networks under optimal conditions, a degraded channel can amplify the effects of traditional attacks.

\end{takeaway}

\section{Impact on Security Measures}
\label{sec:ids}

As the introduction of the 5G can amplify the effects of network attacks and degrade the security of the system, the question arises whether traditionally deployed countermeasures remain effective.
A vital part of any defense-in-depth strategy in industrial systems are anomaly-based intrusion detection systems, which aim to detect attacks before they cause damage to the physical process~\cite{wolsing2022ipal}.
These systems work under the assumptions that the behavior of the physical process and communication patterns are predictable, while attacks change this behavior noticeably.

However, due to the fluctuating and unreliable nature of the wireless medium, these assumptions may no longer hold.
Therefore, we assess the performance of current, state-of-the-art detection approaches in 5G industrial settings, especially under varying channel conditions.
We base our analysis on two widely researched communication-based detection approaches, which build their models based on network timings and packet sequences (e.g.,~\cite{lin2018IaT,salem2016IaT,ferling2018dtmc, caselli2015sequences}). %
We exclude approaches that solely monitor the process state (e.g.,~\cite{wolsing2022simple, aoudi2018Pasad}) or host-behavior (e.g.,~\cite{doumanidis2023Icsml}) as these systems are unaffected by the fluctuating physical channel, and their effectiveness presents  challenges orthogonal to our work. 
However, we note that an effective intrusion detection strategy should monitor multiple characteristics simultaneously~\cite{ensembleIDS}, rendering the methods we evaluate a vital part of modern cyber-defense.

To assess the performance of network-based intrusion detection mechanisms in 5G-enabled ICS, we evaluate the impact of 5G on model quality (§\ref{sec:ids-model}) and alert behavior (§\ref{sec:ids-alert}) across all deployment scenarios (cf.~§\ref{sec:methodology}) and under dynamic channel conditions.

\subsection{Impact on Timing-based Intrusion Detection}
\label{sec:ids-model}

A widely used strategy of network-based intrusion detection in industrial systems relies on the inter-arrival timing of network packets, i.e., the time that passes between two network packets of the same communication flow or type~\cite{salem2016IaT,lin2018IaT}.
The underlying assumption of this approach is that attacks observably impact packet timings compared to benign conditions.
In the following, we assess whether this assumption still holds when using 5G in \acp{ICS}.

\noindent
\textbf{Experimental Setup.}
To assess the impact of 5G on timing-based intrusion detection, we utilize the IPAL framework~\cite{wolsing2022ipal} which provides an open-source implementation of Lin et al.'s~\cite{lin2018IaT} detection approach.
This approach extracts the inter-arrival times for the individual flows in the network and models them using bounds based on average timings within a given window.
Consequently, to assess the suitability of a timing-based detection approach for \ac{ICS} using 5G, we compare the model quality using the distributions of inter-arrival times for each deployment scenario. 
To this end, we extract attack-free network traffic from \name{} for each deployment as training data and compute the three corresponding models.

\begin{figure}
    \centering
    \includegraphics{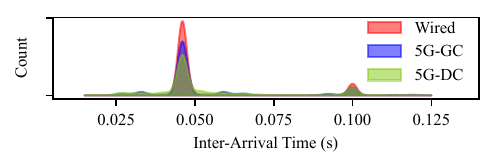}
    \caption{The distributions of inter-arrival times in the {\colorbox{transparentRed}{Wired}}, {\colorbox{transparentBlue}{\goodChannel{}}}, and {\colorbox{transparentGreen}{\badChannel{}}} scenarios showcase that increased jitter widens the bounds of the inter-arrival time detector~\cite{lin2018IaT}, potentially allowing malicious behavior.
    }
    \label{fig:iat-model}
\end{figure}

\begin{figure*}[t]
    \centering
    \includegraphics[width=\textwidth]{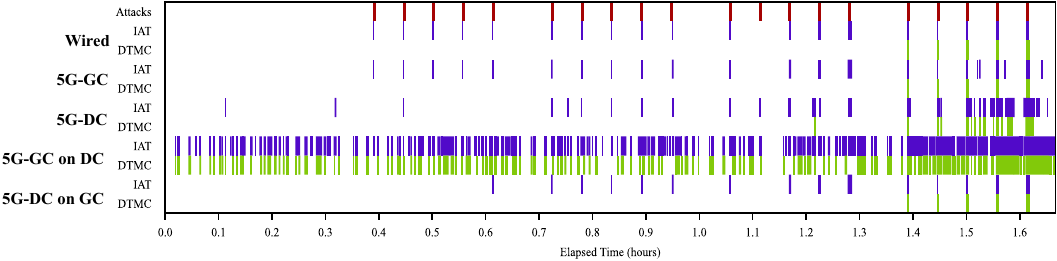}
    \caption{The alert behavior of communication-based detectors deployed in the \emph{Wired} and \emph{\goodChannel{}} deployments show the accuracy of the inter-arrival timing-based detector~\cite{lin2018IaT} (\textcolor{blue}{IaT - blue}) and sequence-based intrusion detection~\cite{ferling2018dtmc} (\textcolor{myGreen}{DTMC - green}). 
    The models derived from the \emph{\badChannel{}} produce additional false alerts and detects fewer attacks due to increased jitter and delays.
    Deploying the 5G models in different channel conditions (\goodChannel{} on DC and \badChannel{} on GC) impacts their alert behavior significantly, even rendering benign behavior indistinguishable from attack periods (5G-GC on DC).}
    \label{fig:alerts}
\end{figure*}

\noindent
\textbf{Results.}
Fig.~\ref{fig:iat-model} shows the distribution of inter-arrival times for all flows for the \emph{Wired}, \emph{\goodChannel{}}, and \emph{\badChannel{}} model. 
We observe two peaks in all three distributions around \SI{0.04}{\s} and  \SI{0.10}{\s}.
These peaks arise because the flows within the network tend to average these specific inter-arrival times.
Traditionally, the timing-based detector relies on tight bounds, derived from low network jitter, to identify deviations caused by attacks.
The \emph{Wired} model ({\colorbox{transparentRed}{red}} in Fig.~\ref{fig:iat-model}) exhibits such tight bounds as indicated by the narrower peaks around \SI{0.04}{\s} and \SI{0.10}{\s}.
Although the model trained on \emph{\goodChannel{}}  data ({\colorbox{transparentBlue}{blue}} in Fig.~\ref{fig:iat-model})  exhibits similarly narrow peaks with a similarly small standard deviation for individual flows, we observe additional peaks around \SI{0.03}{\s}, \SI{0.06}{\s}, and \SI{0.09}{\s} in the model's distribution resulting from additional retransmissions caused by the less reliable wireless channel. 
Finally, the model for the degraded wireless channel (\badChannel{}) ({\colorbox{transparentGreen}{green}} in Fig.~\ref{fig:iat-model}) exhibits broader peaks in the distribution, showing higher deviation in the observed timings, potentially marking more behavior as benign.

These results show that degraded channel conditions negatively impact the training data quality of a timing-based detection system, leading to a more relaxed model that allows broader behavior and may detect less malicious activity, thus decreasing its effectiveness.

\subsection{Impact on Alert Behavior}
\label{sec:ids-alert}

In addition to monitoring timings, communication-based intrusion detection systems utilizing packet sequences present another established method for attack detection in industrial settings (e.g.,~\cite{ferling2018dtmc, caselli2015sequences}).
Similar to the timing-based approach, these systems rely on deterministic traffic to produce consistent packet sequences under benign conditions assuming that attacks alter this sequence.
In the following, we assess the suitability of both timing-based and sequence-based mechanisms---on the example of Ferling et al.'s~\cite{ferling2018dtmc} detection approach based on Discrete Time Markov Chains (DTMCs)---for \ac{ICS} using 5G.
Furthermore, we also test how these approaches adapt to a fluctuating wireless medium.

\noindent
\textbf{Experimental Setup.}
We generate a dataset from the network traffic of each deployment and attack scenario (cf.\ §\ref{sec:attacks}) realized with \name{}~\cite{swics}.
Then, we train and deploy each intrusion detection model on each respective dataset to assess detected attacks and false alerts.
Additionally, we evaluate the robustness of the inter-arrival time based and sequence-based detectors under fluctuating channel conditions and thus their suitability for usage in 5G-enabled \acp{ICS}.
To this end, we study alert behavior when models are trained on data from one 5G scenario and deployed in the other (e.g., trained in \goodChannel{} and deployed in \badChannel{}, and vice versa).
Furthermore, we conduct these experiments using the IPAL framework~\cite{wolsing2022ipal}, to allow for a direct comparison between the detectors. %

\noindent
\textbf{Results.}
Fig.~\ref{fig:alerts} compares the ground truth of attacks (top row; {\colorbox{red}{red boxes}) and the alerts of the intrusion detection systems for each channel configuration (i.e., \emph{Wired}, \emph{\goodChannel{}}, \emph{\badChannel{}}) as well as dynamic noise levels simulated by switching training and test data (\emph{\goodChannel{} on DC} and \emph{\badChannel{} on GC}).
Setting a baseline for comparison, the timing-based detector performs nearly perfectly in the \emph{Wired} scenario, detecting 19 out of 20 attacks without raising false alerts.
This performance showcases how the models tight bounds result in high detection capabilities.
Similarly, the \emph{\goodChannel{}} model detects 19 out of 20 attacks in the \goodChannel{} dataset (Fig.~\ref{fig:alerts} \goodChannel{}).
Furthermore, the timing-based detector produces 4 false alerts towards the end of the dataset, which result from jitter in the wireless medium.
DTMC accurately detects 5 out of 20 attacks in the \emph{Wired} and \emph{\goodChannel} deployments, but misses the remaining 15, likely caused by DTMC’s permissiveness toward minor packet sequence changes.

The results for the \emph{\badChannel{}} deployments show a degradation in detection performance for both detection approaches.
The timing-based model trained and evaluated on the degraded channel (Fig.~\ref{fig:alerts} \badChannel{}) still detects 15 out of the 20 attacks, but produces 17 false alerts.
Because most of these false alerts cluster near the end of the dataset, they make the last four true alerts unrecognizable.
Moreover, both approaches produce additional false alerts in the \emph{\badChannel} scenario, caused by high jitter in the noisy channel.

To test the effectiveness of these traditionally well-working detection approaches under the novel challenges presented by the dynamic 5G channel, we evaluate \emph{\goodChannel{}} models on noisy-channel datasets and vice versa.
We observe that the noisy channel alters the inter-arrival times and the packet sequences to such a degree that both \emph{\goodChannel{}} models flag the behavior as alerts.
These constant false alerts render attack-free periods indistinguishable from attacks. 
Conversely, deploying the \emph{\badChannel{}} model in the noiseless channel (Fig.~\ref{fig:alerts} 5G-DC on GC) reduces alerts.
In this case, improved channel conditions reduce false alerts for both detectors, enhancing DTMC's performance to correctly detect 5 out of 20 attacks, similar to the \emph{Wired} and \emph{\goodChannel{}} baselines.
Similarly, the timing-based approach detects 15 out of 20 attack scenarios\textemdash{}still, 4 less than the model trained under these channel conditions.

These results show that communication-based intrusion detection is sensitive to channel variation, hindering the collection of high-quality training data~\cite{wolsing2024deployment, ahmed2020IdsChallenges}.
Moreover, in 5G communication-based detection becomes unreliable, as static models struggle to distinguish poor channel quality from actual attacks.

\begin{takeaway}
    As current assumption of predictable communication behavior no longer hold under dynamic channel conditions, existing intrusion detection systems become unreliable presenting novel concerns including the collection of training data.
\end{takeaway}

\section{Attacks on the Wireless Interface}
\label{sec:wirelessAttacks}

So far, our analysis has focused on the impact of 5G on traditional \ac{ICS} attacks and security mechanisms. 
However, transitioning to a wireless link introduces new vulnerabilities and increases exposure to external threats, which we examine in this section.
Specifically, we show how an \emph{outsider} attacker can extract critical information about the \ac{ICS} (§\ref{sec:spectrum}), which can be leveraged to perform jamming attacks that disrupt system operation (§\ref{sec:jamming}). While such attacks have been studied in other wireless \acp{ICS} (cf.~§\ref{sec:related}), our results indicate that in 5G, MIMO enables significantly more effective and stealthier jamming, potentially even at low transmission power.

\subsection{Passive Reconnaissance}
\label{sec:spectrum}

Extracting communication patterns from a wired \ac{ICS} is usually challenging, as it requires physical or logical network access. With a wireless medium, however, eavesdropping becomes effortless. 
Anyone with commercial-grade radio equipment, such as an SDR, and open-source software can analyze the wireless spectrum.
Spectrum analyzers are tools that provide a visual representation of signal frequencies, allowing detailed observation of transmission presence, strength, and patterns over time. This enables an attacker to passively monitor communication, identify active frequency bands, and infer key operational characteristics of the system.

To assess susceptibility to passive reconnaissance attacks, we deploy a customized spectrum analyzer module \emph{outside the facility perimeter}. 
Fig.~\ref{fig:spectrogram} shows the monitored 5G channel, where transmission intervals appear as regions of high signal power. 
These observations allow an attacker to infer the \ac{ICS} process cycle and mount targeted reactive jamming attacks~\cite{jamm}, selectively disrupting critical signals\textemdash{}even if traffic is encrypted.

\begin{figure}
    \centering
    \includegraphics{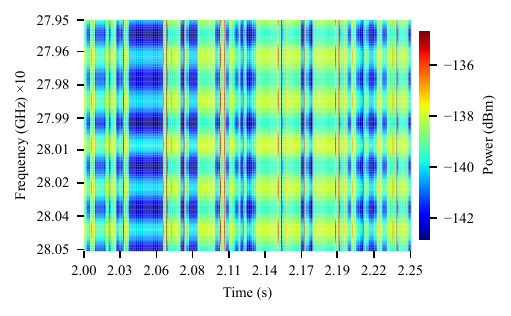}
    \caption{By passively monitoring the 5G channel even from outside the industrial premises, an attacker can infer critical details about an \ac{ICS} such as the process cycle.}
    \label{fig:spectrogram}
\end{figure}

\subsection{Jamming to Disrupt the Physical Process}
\label{sec:jamming}

In 5G networks, DoS attacks can be launched by any adversary within communication range, without requiring network access. 
A jammer interferes with legitimate transmissions by emitting radio signals, potentially degrading or blocking communication. 
In 5G-enabled \acp{ICS}, this is particularly critical as jamming directly targets system availability~\cite{michaelides2025industry5G}. 
We distinguish between constant jamming, which continuously emits noise, and reactive jamming, which transmits only upon detecting legitimate signals (e.g., via spectrum analysis as in §\ref{sec:spectrum}), making it more stealthy.

\noindent
 \textbf{Experimental Setup.} 
 To simulate jamming, we implement a custom physical-layer device in ns-3. When activated, it signals the channel model to inject a high-power signal for a fixed duration, increasing background noise, leading to malformed or dropped packets. 
 We also add MIMO support to reflect real-world mmWave 5G devices (cf.~\ref{sec:background:5g}), enabling disruption via lower-power directional transmissions. 
 We apply jamming intermittently across the process for short intervals, covering about 50\% of the simulation time. 
 This approach lies between reactive and constant jamming. Our aim is to observe its effects at different process stages.

\begin{figure}[t]
    \centering
    \includegraphics[width=1\linewidth]{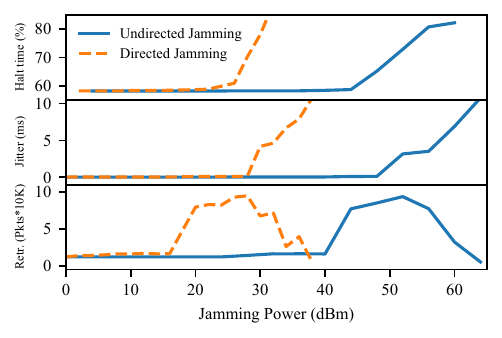}

    \caption{Impact of jamming from outside the factory on the ICS: jitter (Top), retransmissions (Mid), and halting time (Bottom). Directed jamming causes similar disruption to undirected jamming with half the transmission power.}
    \label{fig:jamming}
\end{figure}

\noindent
\textbf{Impact on Network Performance.}  To assess the impact on network performance, we consider two key metrics: jitter (variance in latency) and number of retransmissions. 
In our experiments, we jam until jitter reaches \SI{10}{ms}, a commonly accepted upper bound for time‑critical industrial applications~\cite{michaelides2025industry5G}. Even if average latency remains below \SI{1}{ms}, exceeding this jitter threshold violates latency guarantees. 
As shown in Fig.~\ref{fig:jamming}~(Top), a directed jammer outside the facility using 4$\times$4 antennas can significantly increase jitter with transmission power only slightly above that of a legitimate \ac{UE}, making detection difficult. 
In contrast, undirected jamming requires roughly twice the gNB's transmission power and is therefore easier to detect.
Retransmissions (Fig.~\ref{fig:jamming}~(Mid)) show a similar trend: directed jamming disrupts communication with less than half the power required for undirected jamming. Retransmissions initially increase as packet corruption triggers recovery mechanisms, but eventually decrease once noise prevents successful transmission altogether, effectively collapsing communication. 
This packet loss can be catastrophic\textemdash{}e.g., if safety-critical commands fail to arrive. 
Overall, directed jamming below \SI{20}{dBm} can induce severe packet loss, highlighting that an attacker with a compromised \ac{UE}/gNB and a small antenna array can launch effective, low-power jamming attacks that are difficult to detect and significantly increase risk.

\noindent
\textbf{Impact on the Physical Process.} 
To evaluate the impact of jamming on the physical process, we consider the fraction of time the conveyor belt is stationary.
Under normal conditions, the conveyor belt is stationary roughly 55\% of the time, as it only operates to position a new bottle under the valve.
As shown in Fig.~\ref{fig:jamming} (Bottom), with increased jamming power, the halting time rises because more packets are lost, triggering the safety mechanism (cf.\ §\ref{sec:swics}) that causes a full shutdown when no messages are received within a specified time frame.
Consistent with the impact on network performance, directed jamming causes the same disruption as undirected jamming but requires only about half the transmission power. 
Specifically, directed jamming achieves an 80\% halting time at around \SI{30}{dBm} transmission power, while undirected jamming requires approximately \SI{55}{dBm} to produce the same effect.

\begin{takeaway}
    Introducing a wireless medium to \acp{ICS} broadens the attack surface, enabling stealthy reconnaissance and jamming. In 5G mmWave MIMO environments, an attacker with a compromised device and spectrum analyzer can perform directional jamming to disrupt the ICS without triggering alarms.
\end{takeaway}

\section{Discussion and Future Work}
\label{sec:discussion}

Our analysis shows that 5G communication can amplify traditional \ac{ICS} attacks, challenge existing security measures, and introduce new attack vectors. 
To contextualize these findings, we discuss key methodological choices and limitations, and identify promising countermeasures. 
For each aspect, we also outline directions for future research.

\noindent
\textbf{Choice of Attack Vectors.}
Our work focuses on assessing the security challenges arising from introducing 5G communication into \acp{ICS}.
However, we do not simulate a full 5G protocol stack in \name{}, but instead focus on the impact of the wireless medium---specifically the mmWave spectrum---on the security of the physical process.
Although, we consider vulnerabilities in the 5G infrastructure in our threat model (§\ref{sec:methodology}), the security and weaknesses of 5G protocols and components are generalizable across use-cases and have been extensively studied~\cite{299752,10.1145/3448300.3467817, 10.1145/3734477.3734725,10.1145/3448300.3467826,hussain2019privacy}. 
However, it is important to note that these vulnerabilities expand the traditional attack surface of \acp{ICS}, especially w.r.t. to availability(c.f.~§\ref{sec:impactPhy}).
Future work should therefore focus on integrating a full 5G protocol stack in ns-3 and conducting a 5G-focused security analysis in \acp{ICS}.

\noindent
\textbf{Selection of Intrusion Detection Systems.}
Intrusion detection systems have long been a popular security control for \ac{ICS} due to their ability to retrofit security into legacy systems~\cite{wolsing2022ipal}. 
For our analysis, we focus on two relatively simple intrusion detection strategies that rely on predictable communication patterns, such as packet sequences and inter-arrival times\textemdash{}techniques traditionally effective in ICS~\cite{wolsing2022ipal}. 
Our results show that these approaches struggle under the inherently dynamic conditions of 5G networks, particularly when channel conditions degrade compared to the training phase. 
These findings highlight that traditional ICS security controls cannot be assumed to remain effective in wireless environments and must be re-evaluated and updated.
Operators must either include varying channel conditions in the training data\textemdash{}complicating the already challenging task of collecting representative data~\cite{conti2021survey,wolsing2024deployment, ahmed2020IdsChallenges}\textemdash{}or find alternative protection mechanisms, which can be developed, studied, and evaluated using \name{}.

\noindent
\textbf{Transferability to real ICS.}
We emphasize that behavior of real-world ICSs depends on many factors such as, the specific topology (network or other devices), network setup, device types, and more. 
Therefore, transferring the results presented in this paper might not transferable to other systems ``1-to-1''.
Instead, this study aims to provide general insight on the effects of 5G on ICS security, where we took multiple steps to ensure a close representation of real-life systems.
Thus, to recreate a realistic 5G channel, we utilized network and channel models that are standardized by the 3GPP and that are calibrated against real-world 5G networks.
Furthermore, we conduct our experiments using 5G mmWave and thus ensure that any observed effects are likely to be amplified in other wireless technologies commonly used in industry that do not offer such high performance guarantees.
At last, we intentionally select a simple and robust bottle-filling process with the rationale that if attacks can cause substantial disruptions even in such a simple and robust scenario, their effects are likely to be even larger in more complex systems as they would occur in the real-world.
In total, these decisions aim to provide a realistic understanding of ICS security under 5G.
Nevertheless, \name{} allows replacing the underlying ICS or
the wireless technology with minimal effort to conduct research on more involved deployments in the future.

\noindent
\textbf{Countermeasures.} The choice of countermeasures in industrial 5G networks is strongly influenced by the requirements of the underlying physical process. 
A key advantage of 5G is its support for optional user plane security mechanisms, which provide encryption and integrity protection across the 5G infrastructure~\cite{michaelides2025industry5G}. These controls can be used in networks with latency requirements between 1-10 ms. However, for sub-ms latency requirements, such mechanisms may become impractical due to their overhead. 
In such cases  security protocols such as TLS or IPsec can be employed~\cite{MichaelideNLS,Dekker2020PerformanceCO}, as they provide end-to-end  protection avoiding additional per-hop processing delays.
Further, latency-free alternative controls—such as intrusion detection systems that account for varying channel conditions or approaches based on payload inspection—can be employed to further enhance system security.

To mitigate attacks on the wireless interface, techniques such as beamforming, frequency hopping, and MIMO-based spatial filtering have been proposed~\cite{SKOKOWSKI2022258}. 
Nevertheless, complete protection is unattainable without  enclosing the factory in a Faraday cage, making physical security measures\textemdash{}e.g., extending the controlled physical perimeter to increase the cost and detectability of jamming\textemdash{}an essential complement to cybersecurity. 

\noindent\textbf{Ethical Considerations.} 
Analyzing attacks on \acp{ICS} raises ethical considerations. To minimize harm, we rely on a fully virtual testbed, eliminating risks to deployed systems and human operators. All attacks are implemented in an abstracted simulation, do not produce executable malicious code, and are based solely on publicly available information. This enables us to study the security implications of 5G in \acp{ICS} while reducing the risk of misuse.

\section{Conclusion}

To address the increasing demand for flexibility in modern \acp{ICS}, a transition from wired to 5G communication is underway~\cite{aijaz2020private5G, michaelides2025industry5G, seferagic202WirelessSurvey}.
However, the security implications of switching to 5G in \acp{ICS} remain underexplored.
To fill this gap, we present \name{}~\cite{swics}, the first virtual ICS-security testbed utilizing 5G and capturing a thoroughly modeled physical process.
To enable reproducibility and facilitate further research, we make \name{} freely accessible~\cite{swics}.

Using \name{}, we assess the impact of 5G on the security of \ac{ICS}, including the stability of the physical process under various attack scenarios~(§\ref{sec:impactPhy}), the consequences for existing security measures (§\ref{sec:ids}), and the increased risk from novel wireless-only attack vectors~(§\ref{sec:wirelessAttacks}).
To this end, we directly compare traditional wired communication with 5G under varying channel conditions.
Our findings show that under optimal conditions 5G provides comparable stability and security as the wired medium.
However, a degraded channel poses substantial challenges for the security of \acp{ICS}, especially w.r.t.\ the susceptibility to cyber-attacks as well as communication-pattern based attack detection.
Lastly, the wireless interface of 5G opens novel attack vectors for reconnaissance and jamming attacks, which may compromise availability.

Our results highlight the need to adapt existing \ac{ICS} security measures when transitioning to 5G to account for fluctuations in channel quality and to consider resilience strategies against wireless-attacks such as jamming.
Ultimately, we emphasize the need for comprehensive countermeasures fit for  ultra-low latency applications that explicitly account for the effects of a 5G channel.

\begin{acks}
  We sincerely thank the shepherd and the reviewers for their feedback on improving this work.
  Funded by the Foundation for Innovation in Higher Education (Freiraum Project RealistICS) and the German Federal Ministry of Research, Technology and Space (BMFTR) under funding reference number 16KIS2409K (6GEM+).
  The authors are responsible for the content of this publication.
\end{acks}

\bibliographystyle{ACM-Reference-Format-limit}
\bibliography{references.bib}

\end{document}